%% file: draft.tex
\documentclass[traditabstract]{aa}

\usepackage				{graphicx}
\usepackage                             {amsmath}
\usepackage				{txfonts}
\usepackage				{natbib}
\bibpunct{(}{)}{;}{a}{}{,} 

\usepackage{epstopdf}

\begin{document}
   \title{Constraints on Cosmological Parameters from Strong Gravitational Lensing by Galaxy Clusters}

   \author{B. Zieser
          \and
          M. Bartelmann
          }

   \institute{Zentrum f\"ur Astronomie, Institut f\"ur Theoretische Astrophysik, Albert-Ueberle-Str. 2, 69120 Heidelberg, Germany}

  \date{Manuscript version April 2, 2012}

  \abstract
   {We investigate how observations of strong lensing can be used to infer cosmological parameters, in particular the equation of state of dark energy. We focus on the growth of the critical lines of lensing clusters with the source redshift as this behaviour depends on the distance-redshift relation and is therefore cosmologically sensitive.
   Purely analytical approaches are generally insufficient because they rely on axisymmetric mass distributions and thus cannot take irregular critical curves into account. We devise a numerical method based on the Metropolis-Hastings algorithm: an elliptical generalization of the NFW density profile is used to fit a lens model to an observed configuration of giant luminous arcs while simultaneously optimizing the geometry. A semi-analytic method, which derives geometric parameters from critical points, is discussed as a faster alternative. We test the approaches on mock observations of gravitational lensing by a numerically simulated cluster.
   We find that no constraints can be derived from observations of individual clusters if no knowledge of the underlying mass distribution is assumed. Uncertainties are improved if a fixed lens model is used for a purely geometrical optimization, but the choice of a parametric model may produce strong biases.}


   \maketitle
%

\input{sec1}

\input{sec2}

\input{sec3}

\input{sec4}

\input{sec5}

\input{sec6}

\bibliographystyle{aa}
\bibliography{references}

\end{document}

%% file: sec1.tex
\section{Introduction}
In the widely accepted 'concordance model' of cosmology a cosmological constant accounts for more than 70 per cent of the overall energy density in the universe. It is considered an important feature mainly because it can account for both the spatial flatness and the accelerated expansion of the universe. Still there is no physical explanation that is generally considered satisfying, and the observational evidence can be reproduced in models that instead introduce dark energy, characterized by negative pressure.
Observations of gravitational lensing may help determine its equation of state. Lensing phenomena are sensitive to the geometry of the cosmological background since the appearance of an image depends on the distances between source, lens and observer. If we can obtain information about these distances from observations, we can relate them to redshifts to constrain spacetime curvature, which is governed by cosmological parameters.

In this paper we study strong gravitational lensing. Since a mass distribution needs to feature very high densities to act as a strong lens, galaxy clusters are suitable subjects of investigation. The critical lines of such lenses grow with the redshift of the source. The methods discussed here aim to infer geometrical information from observations of giant arcs, which trace the critical lines, and consequently constrain the equation of state of dark energy. An exploratory study of the concept was presented by \citet{2005MNRAS.362.1301M}. Other authors followed similar approaches to analyse individual clusters, e. g. \citet{2002A&A...393..757S}, \citet{2004A&A...417L..33S}, \citet{2009MNRAS.396..354G}. Moreover weak lensing can be studied with the same goal, as presented by \citet{2011MNRAS.414.1840M}, for instance.

We give a short summary of the underlying theory in Sect. \ref{sec:theory}. In Sect. \ref{sec:analytic} we present analytic studies of cluster lensing. While they are restricted to axisymmetric lenses, they can provide us with estimates of the influence of the dark energy equation of state on strong lensing features. In Sect. \ref{sec:mcmc} we turn to numerical approaches and develop Markov chain Monte Carlo methods that aim to fit a set of parameters, characterizing the lens model and the geometry, to an observed image configuration. In Sect. \ref{sec:semi} an alternative approach is presented, which infers the geometry from the observed scaling of the critical lines at different redshifts. Finally in Sect. \ref{sec:conclusions} we discuss the performance of the various approaches and point out possible sources of errors.

%% file: sec2.tex
\section{Theoretical background}\label{sec:theory}
\subsection{Cosmological model}
We assume that the universe is spatially flat and characterized by the Friedmann-Lema\^itre-Robertson-Walker metric
\begin{equation}
 \mathrm{d}s^2=-c^2\mathrm{d}t^2+a^2(t)\left[\mathrm{d}w^2+w^2\left(\mathrm{d}\theta^2+\sin^2\theta\,\mathrm{d}\phi^2\right)\right],
\end{equation}
where $t$ denotes the coordinate time, $w$ is the comoving radial coordinate and $\theta$ and $\phi$ are the azimuthal and polar angles. The evolution of the scale factor $a(t)$ is governed by the Friedmann equations,
\begin{equation}\label{eq:Friedmann1}
 \left(\frac{\dot{a}}{a}\right)^2=\frac{8\pi G}{3}\rho
\end{equation}
and
\begin{equation}\label{eq:Friedmann2}
 \frac{\ddot{a}}{a}=-\frac{4\pi G}{3}\left(\rho+\frac{3p}{c^2}\right).
\end{equation}
We have omitted the curvature terms as well as those involving a cosmological constant. Instead we consider a dark energy component with the equation of state $p_{\mathrm{x}}=w\rho_{\mathrm{x}}c^2$, such that its density is given by
\begin{equation}
 \rho_{\mathrm{x}}(a)=\rho_{\mathrm{x},0}\,\exp\left(-3\int_1^a\,\frac{1+w(a')}{a'}\,\mathrm{d}a'\right).
\end{equation}
To drive the expansion of the universe $w<-1/3$ is required. From Eq. \eqref{eq:Friedmann1} the Hubble function $H=\dot{a}/a$ can be derived:
\begin{equation}\label{eq:Hubble}
H(z)=H_0\sqrt{\Omega_{\mathrm{m},0}(1+z)^3+\left(1-\Omega_{\mathrm{m},0}\right)\exp\left(3\int_0^z\,\frac{1+w(z')}{1+z'}\,\mathrm{d}z'\right)}.
\end{equation}
Here the matter density parameter $\Omega_{\mathrm{m},0}$ has been introduced. Since curvature is assumed to vanish and the radiation density is negligible in the epochs probed by gravitational lensing, the dark energy density parameter is $\Omega_{\mathrm{x},0}=1-\Omega_{\mathrm{m},0}$.
\subsection{Gravitational lensing}
We summarise only the key aspects of gravitational lensing here. A comprehensive review of the theory was provided by \citet{2010CQGra..27w3001B}.

The relation between the angular positions of an image, $\boldsymbol{\theta}$, and its source, $\boldsymbol{\beta}$, is given by the lens equation:
\begin{equation}
 \boldsymbol{\beta}=\boldsymbol{\theta}-\boldsymbol{\alpha}(\boldsymbol{\theta}).
\end{equation}
$\boldsymbol{\alpha}$ is called the reduced deflection angle and can be written as
\begin{equation}
 \boldsymbol{\alpha}(\boldsymbol{\theta})=\frac{D_{\mathrm{ds}}}{D_{\mathrm{s}}}\boldsymbol{\hat{\alpha}}(\boldsymbol{\theta}),
\end{equation}
where the deflection angle $\boldsymbol{\hat{\alpha}}$ depends only on the lensing mass distribution. $D_{\mathrm{ds}}$ is the distance from the deflector to the source and $D_{\mathrm{s}}$ the distance from the observer to the source. Both are angular diameter distances, with the distance to redshift $z$ given by the integral
\begin{equation}
 D_{\mathrm{A}}\left(z\right)=\frac{c}{1+z}\int_0^z\frac{\mathrm{d}z'}{H\left(z'\right)}.
\end{equation}
Reduced deflection angles thus depend on the lens and source redshifts.

In the thin-screen approximation the lensing potential $\psi$ is defined as the Newtonian potential projected on to the lens plane,
such that its gradient is the reduced deflection angle, $\boldsymbol{\alpha}(\boldsymbol{\theta})=\boldsymbol{\nabla}_{\theta}\psi(\boldsymbol{\theta})$. Image distortions are quantified by the Jacobian $J$ of the lens mapping ($J_{\mathrm{ij}}=\partial\beta_{\mathrm{i}}/\partial\theta_{\mathrm{j}}$), specifically by the convergence
\begin{equation}
 \kappa\left(\boldsymbol{\theta}\right)=\frac{1}{2}\left(\frac{\partial^2\psi}{\partial\theta_1^2}+\frac{\partial^2\psi}{\partial\theta_2^2}\right)
\end{equation}
and the shear
\begin{equation}
 \gamma_1\left(\boldsymbol{\theta}\right)=\frac{1}{2}\left(\frac{\partial^2\psi}{\partial\theta_1^2}-\frac{\partial^2\psi}{\partial\theta_2^2}\right), \quad \gamma_2\left(\boldsymbol{\theta}\right)=\frac{\partial^2\psi}{\partial\theta_1\partial\theta_2}.
\end{equation}
The convergence is proportional to the Laplacian of the lensing potential, $\boldsymbol{\nabla}^2\psi(\boldsymbol{\theta})=2\kappa(\boldsymbol{\theta})$. It measures the projected surface mass density $\Sigma$ of the lens in units of the critical density:
\begin{equation}
 \kappa(\theta)=\frac{\Sigma(\theta)}{\Sigma_{\mathrm{cr}}},\quad \Sigma_{\mathrm{cr}}=\frac{c^2}{4\pi G}\frac{D_{\mathrm{s}}}{D_{\mathrm{d}}D_{\mathrm{ds}}}
\end{equation}
Critical lines are formed by those points in the lens plane where the lens mapping is singular, $\det J=(1-\kappa)^2-\gamma_1^2-\gamma_2^2=0$. As the lensing potential changes with the source redshift, so does the shape of the critical curves.

%% file: sec3.tex
\section{Analytical approaches}\label{sec:analytic}
\subsection{The singular isothermal sphere}
\begin{figure}
\resizebox{\hsize}{!}{\includegraphics{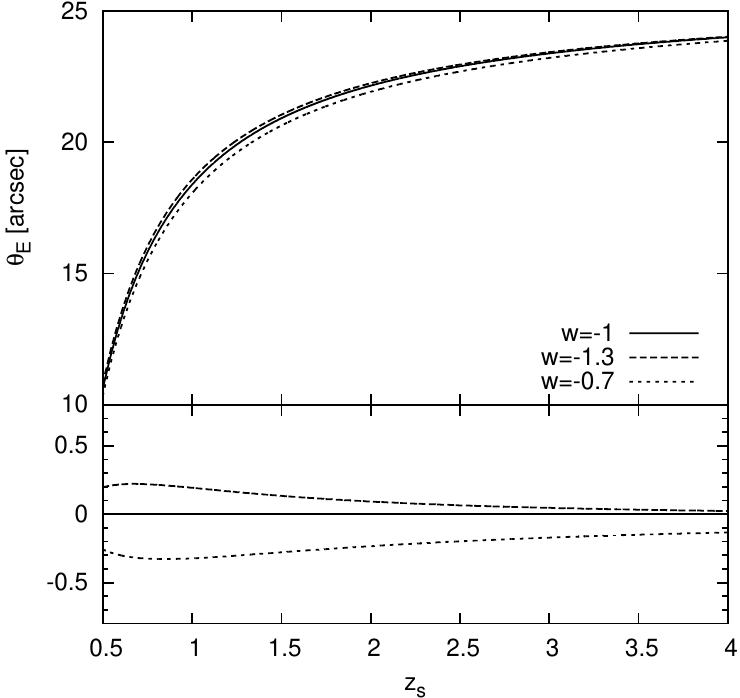}}
\caption{Growth of the Einstein radius $\theta_{\mathrm{E}}$ of a singular isothermal sphere ($\sigma=1000\,\mathrm{km}\,\mathrm{s}^{-1}$) with the source redshift $z_{\mathrm{s}}$ for different values of the equation of state parameter $w$. In the lower panel the deviations from $\Lambda$CDM are shown.}
\label{fig1}
\end{figure}

We first examine the lensing behaviour of the singular isothermal sphere profile. In this model the density of a galaxy cluster is described by the function
\begin{equation}\label{eq:sis}
 \rho(r)=\frac{\sigma^2}{2\pi Gr^2},
\end{equation}
where $\sigma$ is the velocity dispersion of the cluster members. Such a lens is characterised by the Einstein radius
\begin{equation}\label{eq:eradius}
 \theta_{\mathrm{E}}=4\pi\left(\frac{\sigma}{c}\right)^2\frac{D_{\mathrm{ds}}}{D_{\mathrm{s}}}.
\end{equation}
Figure \ref{fig1} shows the dependence of the Einstein radius on the source redshift in different dark energy cosmologies for a cluster with velocity dispersion $\sigma=1000\,\mathrm{km}\,\mathrm{s}^{-1}$ at redshift $z_{\mathrm{d}}=0.3$. In addition to the $\Lambda$CDM scenario $w=-1$, we choose to examine the behaviour for a constant equation of state parameter $w=-0.7$ and an extreme case of dark energy, a phantom model with $w=-1.3$. While the Einstein radius itself takes values of up to $25\arcsec$ for the redshift range considered, differences between the cosmologies are less than $1\arcsec$ and largest if the source is located close to the cluster.

\begin{figure}
\resizebox{\hsize}{!}{\includegraphics{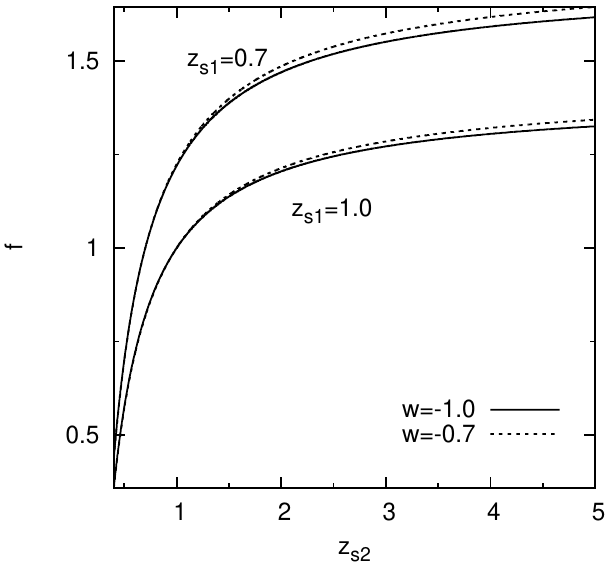}}
\caption{Growth of the geometry factor $f$ with the source redshift $z_{\mathrm{s2}}$ for two different dark energy cosmologies. The fixed reference redshift is $z_{\mathrm{s1}}=1.0$ or $z_{\mathrm{s1}}=0.7$ respectively.}
\label{fig2}
\end{figure}

To obtain the angular diameter distance ratio $D_{\mathrm{ds}}/D_{\mathrm{s}}$ from a single measured Einstein radius $\theta_{\mathrm{E}}$ the velocity dispersion $\sigma$ of the lens must be known. We can eliminate it from our analysis if we study the growth of the critical curve with the source redshift, comparing two Einstein radii $\theta_{\mathrm{E1}}$, $\theta_{\mathrm{E2}}$ at different redshifts $z_{\mathrm{s1}}$, $z_{\mathrm{s2}}$. The resulting ratio
\begin{equation}\label{eq:f}
 f=\frac{\theta_{\mathrm{E2}}}{\theta_{\mathrm{E1}}}=\frac{D_{\mathrm{ds2}}}{D_{\mathrm{s2}}}\frac{D_{\mathrm{s1}}}{D_{\mathrm{ds1}}}
\end{equation}
appears frequently in our analyses, and we label it the \textit{geometry factor}. Via this factor, cosmological parameters determine lensing properties of an object. It should be noted that the proportionality between the geometry factor and the Einstein radius is peculiar to the SIS model. For a fixed redshift $z_{\mathrm{s1}}$, the geometry factor increases with the redshift $z_{\mathrm{s2}}$ of the second source, rising steeply behind the lens and flattening out for high redshifts. In Fig. \ref{fig2} this behaviour is displayed for a lens at redshift $z_{\mathrm{d}}=0.3$ and fixed sources at $z_{\mathrm{s1}}=1.0$ or $z_{\mathrm{s1}}=0.7$ respectively. Differences between the two cosmologies shown are most prominent if the two sources are at a high distance from each other, since this configuration corresponds to a large 'lever arm'. For a cosmological analysis it is therefore the most convenient to study pairs of arcs in which one source lies closely behind the lens and the other at a much higher redshift.

\subsection{The NFW profile}
The NFW profile
\begin{equation}
 \rho(r)=\frac{\rho_{\mathrm{s}}r_{\mathrm{s}}^3}{r(r_{\mathrm{s}}+r)^2}.
\end{equation}
is arguably the most commonly used parametric model for the matter distribution of dark matter haloes. For its Einstein radius no analytic expression exists, but \citet{1996A&A...313..697B} provided the convergence:
\begin{equation}\label{eq:kappa}
 \kappa(x)=\frac{2\kappa_{\mathrm{s}}}{x^2-1}\left(1-\frac{2}{\sqrt{1-x^2}}\mathrm{arctanh}\sqrt{\frac{1-x}{1+x}}\right).
\end{equation}
$x=r/r_{\mathrm{s}}$ is the dimensionless radial coordinate in the lens plane and $\kappa_{\mathrm{s}}=\rho_{\mathrm{s}}r_{\mathrm{s}}\Sigma_{\mathrm{cr}}^{-1}$; note that Eq. \eqref{eq:kappa} is valid only for $x<1$, i. e. inside the scale radius, which encompasses the strong lensing region. We start our analytic approach from a property of axisymmetric lenses, namely the condition that the mean convergence inside the tangential critical line of radius $\theta_{\mathrm{E}}$ has to equal unity:
\begin{align}\label{eq:mean-kappa}
 1&=\frac{1}{\pi\theta_{\mathrm{E}}^2}\int_0^{\theta_{\mathrm{E}}}2\pi\theta\,\mathrm{d}\theta\,\kappa(\theta).
\end{align}
If we set up this equation for two different source redshifts $z_{\mathrm{s1}}$ and $z_{\mathrm{s2}}$, rearrange and divide the two equations, we arrive at the form
\begin{equation}\label{eq:lhs-rhs}
 \frac{D_{\mathrm{ds1}}}{D_{\mathrm{s1}}}\frac{D_{\mathrm{s2}}}{D_{\mathrm{ds2}}}=\frac{\theta_{\mathrm{E1}}^2}{\theta_{\mathrm{E2}}^2}\frac{g\left(x_{\mathrm{E2}}\right)}{g\left(x_{\mathrm{E1}}\right)}.
\end{equation}
The function $g(x)$ represents the integral
\begin{align}
 g(x)&=2\int_0^xx\,\mathrm{d}x\,\frac{\kappa(x)}{\kappa_{\mathrm{s}}}\notag\\
&=4\ln\frac{x}{2}+\frac{8}{\sqrt{1-x^2}}\mathrm{arctanh}\sqrt{\frac{1-x}{1+x}},
\end{align}
solved by \citet{1996A&A...313..697B}. Again we have used dimensionless coordinates in the lens plane, i. e. $x_{\mathrm{E}1,2}=D_{\mathrm{d}}\theta_{\mathrm{E}1,2}/r_{\mathrm{s}}$.
The left-hand side of Eq. \eqref{eq:lhs-rhs} is a function of the source redshifts, while the right-hand side depends on the Einstein radii for those redshifts; both sides also depend on cosmological parameters since these determine the distance-redshift relation. In this form the equation admits a simple graphic solution: plotting each side against the equation-of-state parameter $w$, the intersection of both curves marks the true value.

To test whether the relation can be exploited that way, we consider as an example a halo of mass $M=1.0\times10^{15}\,h^{-1}\,\mathrm{M}_{\sun}$ and scale radius $r_{\mathrm{s}}=310\,h^{-1}\,\mathrm{kpc}$ at a redshift of $z_{\mathrm{d}}=0.3$ and sources at redshifts $z_{\mathrm{s1}}=0.7$ and $z_{\mathrm{s2}}=3.0$. The Einstein radii for those redshifts are $\theta_{\mathrm{E1}}=7\farcs9$ and $\theta_{\mathrm{E2}}=21\farcs9$ in $\Lambda$CDM cosmology. In reality, we have to resort to estimates of the scale radius, such as best-fitting values, which introduce errors. Einstein radii have to be determined from the positions of observed arcs. While the latter can be expected to trace the critical lines of the cluster, uncertainties in the deduced Einstein radii may be at least as large as their widths and thus of the order of $1\arcsec$.

\begin{figure}
 \centering
  \resizebox{\hsize}{!}{\includegraphics{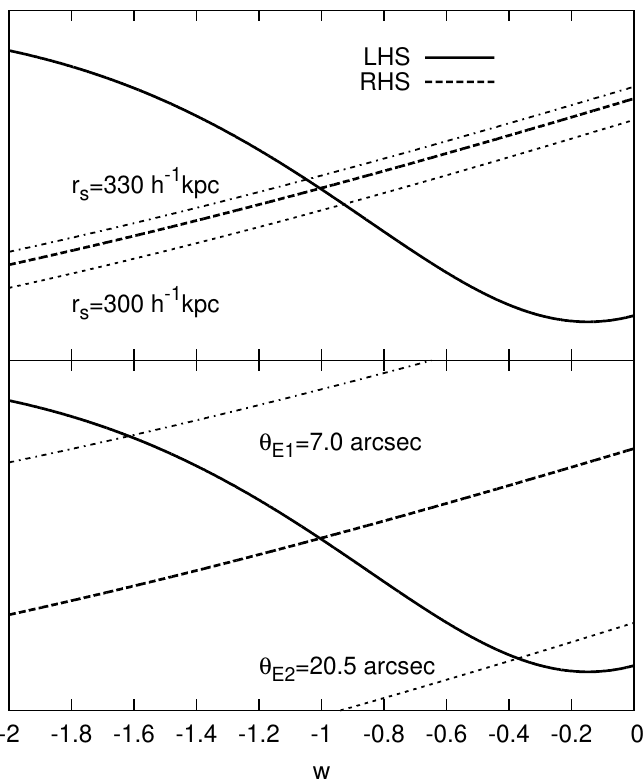}}
  \caption{Left-hand side (LHS) and right-hand side (RHS) of Eq. \eqref{eq:lhs-rhs} plotted against the equation of state parameter $w$ for various estimates of the scale radius $r_{\mathrm{s}}$ (upper panel) and Einstein radii $\theta_{\mathrm{E1}}$, $\theta_{\mathrm{E2}}$ (lower panel). For each curve only the specified quantity is changed from its true value. The thick line in each panel corresponds to the correct values $r_{\mathrm{s}}=310\,h^{-1}\,\mathrm{kpc}$, $\theta_{\mathrm{E1}}=7\farcs9$, $\theta_{\mathrm{E2}}=21\farcs9$.}
  \label{fig3}
\end{figure}

Figure \ref{fig3} shows the influence of such errors on Eq. \eqref{eq:lhs-rhs}. In the first plot, the correct values for the Einstein radii are used, but the estimate for the scale radius is too high or too low respectively, while in the second plot one of the Einstein radii is underestimated. In each calculation, the correct values are used for all but the specified quantity. The left-hand side of the equation is independent of such errors since, as stressed before, it depends only on the source redshifts, of which exact knowledge can safely be assumed here.

Inaccuracies in the scale radius estimate do not have a very large effect on the result for the equation of state parameter, shifting the value of $w$ by $\Delta w\sim0.1$ in this example. The determination of the critical line poses a larger problem, with an error of $\Delta\theta_{\mathrm{E}}\sim1\arcsec$ in the Einstein radius translating into a deviation of $\Delta w\sim0.6$. Another caveat is given by the fact that galaxy clusters often possess significant ellipticity. Based on our analysis so far it seems unlikely that these simple, spherically symmetric profiles provide a sufficient approximation, if only for the fact that it is unclear how  a robust measure of the Einstein radius can be obtained from arc positions in the case of non-circular critical lines.

%% file: sec4.tex
\section{Markov chain Monte Carlo methods}\label{sec:mcmc}
\subsection{The lens model}
To avoid the assumption of axisymmetry, we focus on an elliptical generalization of the NFW profile used by \citet{2006ApJ...642...39C}.
This model is based on six parameters that characterize the cluster constituting the gravitational lens: the coordinates $(x_{\mathrm{c}},y_{\mathrm{c}})$ of its centre, the scale convergence $\kappa_{\mathrm{s}}$, the scale radius $r_{\mathrm{s}}$, the ellipticity $\epsilon$ and the position angle $\phi$. To calculate the deflection angle field the coordinate frame is shifted and rotated in such a way that the cluster centre determines the origin and the coordinate axes coincide with the axes of the ellipse. Then an elliptical radius is introduced,
\begin{equation}
 \rho=\sqrt{x^2(1-\epsilon)+\frac{y^2}{1-\epsilon}}.
\end{equation}
To obtain the deflection field, the NFW lensing potential $\psi$ is evaluated at this radius and $\mathbf{\alpha}$ is calculated by differentiation with respect to the coordinates in the original frame. The convergence $\kappa$ and shear $\gamma$ can be computed by further differentiation.
Deflection angles obtained in this way are valid only for a reference source redshift $z_{\mathrm{r}}$. For each additional source $i$ at a different redshift $z_{\mathrm{si}}$, values have to be multiplied by the geometry factor
\begin{equation}
 f_{\mathrm{i}}=\frac{D_{\mathrm{A}}(z_d,z_{\mathrm{si}})}{D_{\mathrm{A}}(0,z_{\mathrm{si}})}\frac{D_{\mathrm{A}}(0,z_r)}{D_{\mathrm{A}}(z_d,z_r)}.
\end{equation}
Given an image point $\boldsymbol{\theta}$ originating from source $i$, the corresponding source point $\boldsymbol{\beta}$ is located using the lens equation
\begin{equation}
 \boldsymbol{\beta}=\boldsymbol{\theta}-f_{\mathrm{i}}\boldsymbol{\alpha},
\end{equation}
with the appropriately rescaled deflection angle $f_{\mathrm{i}}\boldsymbol{\alpha}$ entering. By scanning a grid on the lens plane for points that fulfil the lens equation for this source position, all image points are located.
\subsection{The chi-square}
To quantify how well a set of parameters describes the observed lensing effects, we follow \citet{2006ApJ...642...39C} in introducing a $\chi^2$-function that has three contributions, $\chi^2=\chi_1^2+\chi_2^2+\chi_3^2$:
\begin{itemize}
\item We demand that the first should measure the extent to which the observed images can be reproduced. Let $N$ be the number of data points with coordinates $(x_{\mathrm{i}},y_{\mathrm{i}})$ and $(u_{\mathrm{j}},v_{\mathrm{j}})$ the coordinates of the predicted image points. For each data point, the closest image point $(u_{\mathrm{cl,i}},v_{\mathrm{cl,i}})$ is identified, leading to
\begin{equation}
 \chi_1^2=\frac{1}{N}\sum_{i=1}^N\,\frac{\left(x_{\mathrm{i}}-u_{\mathrm{cl,i}}\right)^2+\left(y_{\mathrm{i}}-v_{\mathrm{cl,i}}\right)^2}{\sigma_{\mathrm{i}}^2}.
\end{equation}
This form assumes that data points are distributed around the predicted image points in a Gaussian fashion, with a standard deviation of $\sigma_{\mathrm{i}}$.
\item The second contribution should test whether the predicted images match the data. For each predicted image point (out of $M$ overall), the closest data point is identified. The chi-square contribution is then given by
\begin{equation}
 \chi_2^2=\frac{1}{M}\sum_{i=1}^M\,\frac{\left(u_{\mathrm{i}}-x_{\mathrm{cl,i}}\right)^2+\left(v_{\mathrm{i}}-y_{\mathrm{cl,i}}\right)^2}{\sigma_{\mathrm{i}}^2}.
\end{equation}
If the model gives rise to any additional image points far from the data, it affects this term. Obviously this contribution has the same form as the first, but the roles of data and image points are reversed.
\item Finally, the size of the sources is taken into account. For each of the $N_{\mathrm{s}}$ sources, the centre $(\bar{p_{\mathrm{i}}},\bar{q_{\mathrm{i}}})$ is determined by computing the mean of either coordinate. Then the mean squared distance from the centre is calculated, averaging over the $P_{\mathrm{i}}$ points assigned to the $i$-th source. For a tolerated source size $\sigma_{\mathrm{s}}$ we take
\begin{equation}
 \chi_2^2=\frac{1}{N_{\mathrm{s}}}\sum_{i=1}^{N_{\mathrm{s}}}\,\frac{1}{P_{\mathrm{i}}}\sum_{j=1}^{P_{\mathrm{i}}}\,\frac{\left(p_{i,j}-\bar{p_{\mathrm{i}}}\right)^2+\left(q_{i,j}-\bar{q_{\mathrm{i}}}\right)^2}{\sigma_{\mathrm{s}}^2}.
\end{equation}
as the final chi-square contribution. Again it is assumed that the distribution of points belonging to the same source around its centre is Gaussian. This last contribution is in fact crucial because neglecting it would mean that source shapes and sizes are arbitrary -- in that case, any image configuration could easily be reproduced with a lens of zero mass and the points placed in the source plane perfectly matching the distribution in the image plane. Requiring instead that sources are small and compact is therefore a strong constraint.
\end{itemize}
We find that a reasonable choice is $\sigma_{\mathrm{i}}=1\farcs0$ and $\sigma_{\mathrm{s}}=0\farcs5$, emphasizing the source size. Note that the behaviour of the algorithm is determined by the ratio between the two parameters, not their absolute sizes.

\subsection{The Metropolis-Hastings algorithm}
We aim to fit a set of lens parameters and geometry factors to observed arcs. Because of the large number of free parameters (six for the lens model and one for each pair of sources) we avoid using a simplex algorithm to minimize the chi-square. Instead we generate Markov chains according to the Metropolis-Hastings algorithm. Given a set of parameter values $x^{(n)}$, a random point $x^{(n+1)}$ is picked from the parameter space and accepted with the transition probability
\begin{equation}
 t\left(x^{(n)}\rightarrow x^{(n+1)}\right)=\mathrm{min}\left[1,\frac{p\left(x^{(n+1)}\right)}{p\left(x^{(n)}\right)}\frac{q\left(x^{(n)}\rightarrow x^{(n+1)}\right)}{q\left(x^{(n+1)}\rightarrow x^{(n)}\right)}\right].
\end{equation}
$p(x)$ is the target probability distribution. We choose the likelihood $L=N\mathrm{exp}(-\chi^2/2)$ (where $N$ accounts for proper normalization), so that the sample is concentrated on regions of high likelihood. By means of the proposal density $q\left(x\rightarrow y\right)$ the step sizes can be limited or parameter ranges set.

\subsection{The \textsc{Skylens} simulator}
To test the algorithm on mock observations, we use \textsc{SkyLens}, a ray-tracing code presented by \citet{2008A&A...482..403M}. It was used for instance by \citet{2009A&A...500..681M} and \citet{2010A&A...514A..93M}. In short, the program generates a distribution of background galaxies, based on a set of real galaxies decomposed into shapelets. Positions in a specified field of view and orientations are randomly selected. If desired, all sources can be placed at a fixed redshift: this feature is very useful for studies like ours that are based on observing the change of lensing properties with the source redshift. Observational effects are added to the lensed image, including the sky background, photon noise and seeing as well as instrument noise. The lensed images can be convolved with point spread functions, which are available for several telescopes.

We choose to study  a numerically simulated cluster labelled \textit{g1}, taken from a sample of hydrodynamical simulations by \citet{2006MNRAS.373..397S}. It was obtained from a dark matter simulation by \citet{2001MNRAS.328..669Y} and re-simulated with added baryonic effects at a higher mass and spatial resolution using \textsc{Gadget-2} \citep{2005MNRAS.364.1105S}. The cluster has a mass of $M_{200}=1.14\times10^{15}\,h^{-1}\,\mathrm{M}_{\sun}$ and a best-fitting scale radius of $r_{\mathrm{s}}=0.310\,h^{-1}\,\mathrm{Mpc}$. Principal axis ratios are $b/a=0.64$ and $c/a=0.57$ and the orientation of the main axis relative to the coordinate axes of the simulation box is given by the angles $\theta_x=33.3\degr$, $\theta_y=57.4\degr$ and $\theta_z=96.1\degr$. The cosmological parameters used in the simulation are $\Omega_{\Lambda,0}=1-\Omega_{m,0}=0.7$ (with $\Omega_{b,0}=0.04$) and $h=0.7$. Detailed explanations can be found in papers about other studies using these simulations (e. g. \citet{2005MNRAS.364..753D,2005A&A...442..405P}). Deflection angle maps were computed for different projections by \citet{2008A&A...482..403M}, placing the cluster at redshift $z_d=0.2975$.

We created mock observations of the mass distribution projected along the $z$-axis for the Advanced Camera for Surveys (ACS) on HST and obtained 9 giant luminous arcs at redshifts $z=0.7,1.0,2.0,4.0$.

\subsection{Results}
We choose flat prior distributions for the lens parameters, confining parameter values to fixed intervals. If no assumptions at all are made about the lens, it is difficult to obtain information about the geometry due to the degeneracies involved; e. g. trends in the scale convergence and geometry factor can compensate each other to some extent. Moreover, confining the parameters makes the parameter space smaller and accelerates its exploration. For the geometry factors we do not explicitly exclude any region from the beginning and limit only the step size as the algorithm is designed such that the chain should generally move 'in the right direction' regardless of the starting point. In practice, starting values for the geometry factors must still not be set too high unless care is taken to ensure the correct calculation of the tiny likelihoods and their ratios in particular. Generally starting values between 0 and 5 and search radii between 0.2 and 0.5 lead to reasonable burn-in phases and acceptance rates.

\begin{figure}
 \centering
\resizebox{\hsize}{!}{\includegraphics{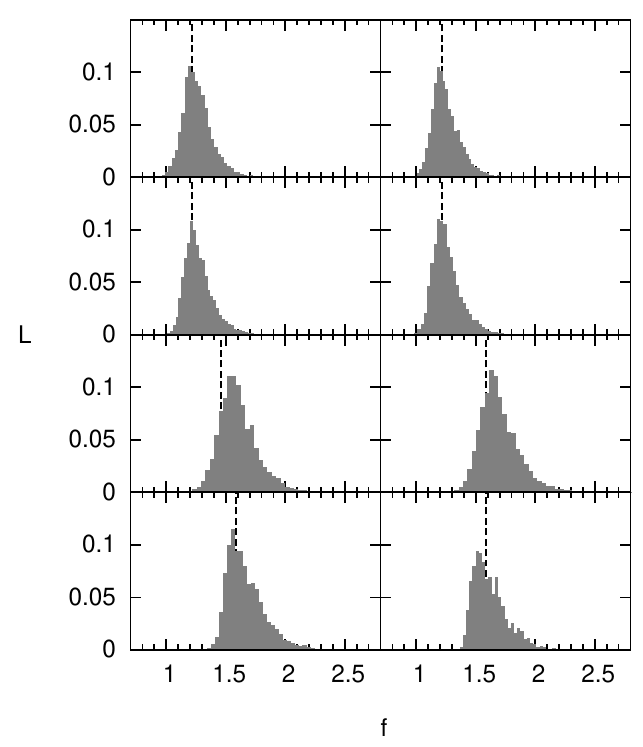}}
  \caption{Distribution of the marginalized likelihood $L$ of the geometry factors $f$ in the full variation; free parameters in each Markov chain are the lens parameters and one geometry factor. The dashed vertical line in each plot indicates the value of the geometry factor in a $\Lambda$CDM cosmology.}
  \label{fig4}
\end{figure}

We consider only one pair of sources at a time: one located at the reference redshift for the lens parameters, the geometry factor of which consequently has the value $1.0$ and is not varied; and a second source at a different redshift. Including the reference redshift helps break the degeneracy between the geometry factor $f$ and the scale convergence $\kappa_{\mathrm{s}}$. Otherwise changes in either quantity can be absorbed in the other, since deflection angles are proportional to the product $f\cdot\kappa_{\mathrm{s}}$. We take the lowest source redshift of $z_{\mathrm{s}}=0.7$ as the reference redshift and follow the procedure described above to produce Markov chains for each geometry factor. From these samples we compute likelihood distributions, marginalizing over the lens parameters. Histograms for the distributions are presented in Fig. \ref{fig4}. In each plot, the true value of the geometry factor, i. e. the value in the $\Lambda$CDM cosmology assumed in the simulation of the images, is marked. The locations of the likelihood peaks generally agree quite well with the true values of the geometry factors. Yet as pointed out in Section \ref{sec:analytic}, geometry factors vary very little between different dark energy cosmologies (cf. Fig. \ref{fig2}).
As $w$ ranges from the rather extreme scenario $w=-2$ to $w=-0.3$ for instance, geometry factors vary by less than 0.1 or roughly 6 per cent (depending on the redshift). On the other hand, the widths of the distributions, which for simplicity we quantify using the standard deviation of the best-fitting Gaussian (despite the skewness), are roughly 8 per cent of their respective mean. The problem lies in the fact that the lens model can easily be adjusted to react to any small change in the geometry. To infer cosmological information, however, a much better 'resolution' is needed.

\begin{figure}
 \centering
\resizebox{\hsize}{!}{\includegraphics{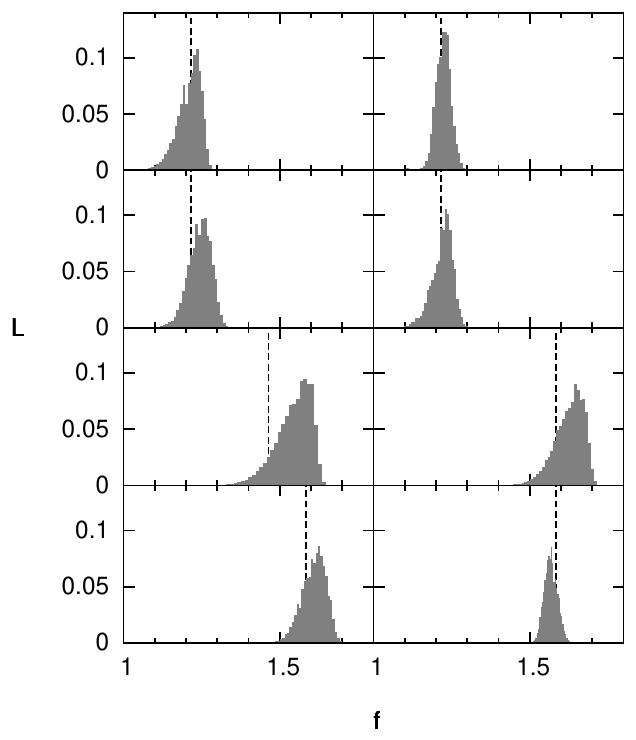}}
  \caption{Distribution of the likelihood $L$ of the geometry factors $f$ in a variation with fixed lens parameters, for a model fitted to 3 arcs at different redshifts. As before the dashed vertical line in each plot indicates the value of the geometry factor in a $\Lambda$CDM cosmology.}
  \label{fig5}
\end{figure}

To address this problem we explore how the method behaves if the lens parameters are kept constant. Ideally, no geometrical assumptions should be made in choosing the lens parameters. In the absence of independent information from effects other than strong lensing, this means that only images at one redshift may be used to fit a model. We attempt this using the code presented by \citet{2006ApJ...642...39C}. However, we find that it is not possible to obtain a reliable parameter set in this way as fit results vary very strongly with different choices of starting values. As we are nonetheless interested in the performance of the algorithm for a fixed lens model, we perform a fit including arcs at three different redshifts.
Again we study the likelihood distributions for the geometry factors, shown in Fig. \ref{fig5}. Compared to the full variation of both lens parameters and geometry factors, the distributions do appear considerably more narrow, with widths of 2-3 per cent of the mean. However, deviations from the $\Lambda$CDM values are still not satisfactory. In most cases, the likelihood peaks are located at geometry factor values that are higher than in $\Lambda$CDM. Comparing the results for several arcs at the same redshift, we note that the likelihood distributions do not seem consistent in that they do not appear to favour the same geometry factors.

In the following we investigate whether our choice of the lens model can account for such deviations.

\subsection{Influence of the lens model}
To check whether the strong biases observed in the likelihood distributions presented above are indeed caused by insufficient knowledge of the lens model, we repeat the procedure using arcs produced by an analytic deflection angle map rather than the simulated cluster. In order to be able to carry out calculations analytically, we define the deflection angle field by a simple power law:
\begin{equation}\label{eq:plaw}
 \alpha(x)=\alpha_0x^p.
\end{equation}
$x=r/r_0$ is the dimensionless radial coordinate in the lens plane; $\alpha$ is also dimensionless and denotes the deflection angle relative to the angle set by the reference scale $r_0$. The latter is arbitrary but has to comply with $\alpha_0=\alpha(x=1)$. The map is therefore characterized by two parameters.

\begin{figure}
 \centering
\resizebox{\hsize}{!}{\includegraphics{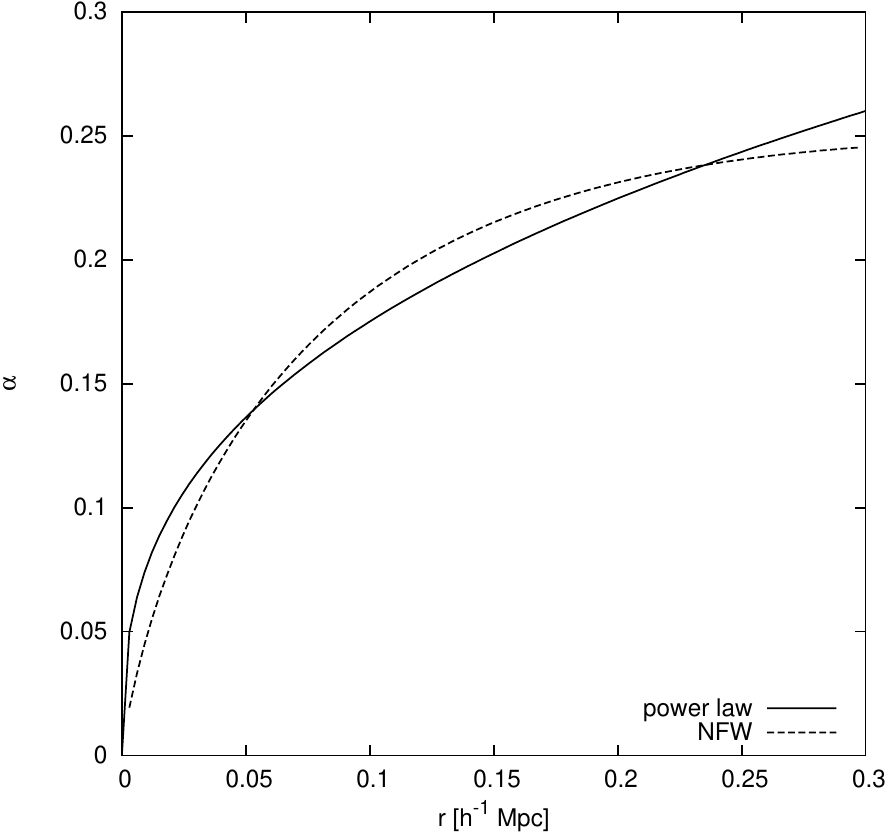}}
  \caption{The dimensionless deflection angle $\alpha$ as a function of the radius $r$ in the lens plane, for a power law $\alpha(r)=0.26[r/(0.3\,h^{-1}\,\mathrm{Mpc})]^{0.36}$ and an NFW halo with scale convergence $\kappa_{\mathrm{s}}=0.2$ and scale radius $r_{\mathrm{s}}=0.3\,h^{-1}\,\mathrm{Mpc}$.}
  \label{fig6}
\end{figure}
We choose a scale of $r_0=0.3\,h^{-1}\,\mathrm{Mpc}$ and set $\alpha_0=0.26$ and $p=0.36$, such that the deflection angle field approximately follows that of an NFW halo of the same scale radius $r_0$ and the scale convergence $\kappa_{\mathrm{s}}=0.2$; the behaviour for both the power law and the specified NFW profile is shown in Fig. \ref{fig6}. We demand that the field should describe the deflection angles for a reference source redshift of $z_{\mathrm{s1}}=0.7$ with the lens located at $z_{\mathrm{d}}=0.3$. In \textsc{SkyLens} simulations, this configuration produces two giant luminous arcs originating from the same source at redshift $z_{\mathrm{s2}}=4.0$.

The MCMC method again provides likelihood distributions for the geometry factor. Since we want to study the influence of the assumed lens model, we run it several times, varying the parameters $\alpha_0$ and $p$. $r_0$ is kept fixed; changing its value has the same effect as changing $\alpha_0$.
\begin{figure}
 \centering
\resizebox{\hsize}{!}{\includegraphics{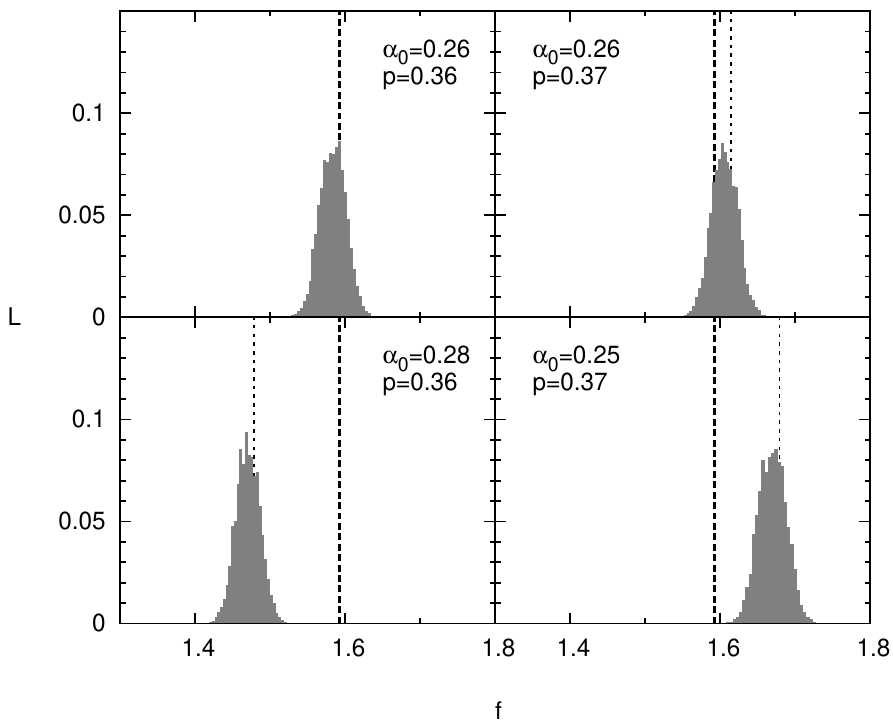}}
  \caption{Likelihood distributions $L$ for the geometry factor $f$ of an arc produced at redshift $z_{s2}=4$ by a power law deflection field. Coefficient $\alpha_0$ and slope $p$ are fixed at the indicated values; true values are $\alpha=0.26$, $p=0.36$. The thick dashed line marks the true geometry factor, thinner lines indicate predictions from analytic calculations for the respective parameter choice.}
  \label{fig7}
\end{figure}
As seen in Fig. \ref{fig7}, the modifications clearly move the likelihood peaks, but they hardly change the shape of the distributions. Based on our exact knowledge of the underlying mass profile in this case we can attempt theoretical predictions:
The deflection field described by Eq. \eqref{eq:plaw} leads to a convergence of
\begin{equation}
 \kappa(x)=\frac{p+1}{2}\alpha_0x^{p-1}
\end{equation}
for the reference redshift; for other redshifts this has to be rescaled by the geometry factor $f$. Now we resort to Eq. \eqref{eq:mean-kappa} again to compute the critical line. The mean convergence inside the radius $x_{\mathrm{c}}$ of the critical line is
\begin{equation}\label{eq:critical}
 1=f\alpha_0x_{\mathrm{c}}^{p-1},
\end{equation}
which leads to an Einstein radius of
\begin{equation}
 x_{\mathrm{c}}=\left(f\alpha_0\right)^{\frac{1}{1-p}}.
\end{equation}
For our choice of parameters, this gives a radius of $x_{c1}=0.122$, corresponding to $11\farcs8$, for the reference redshift $z_{s1}=0.7$. The geometry factor for redshift $z_{s2}=4$ is $f=1.5925$, consequently the Einstein radius should be $x_{c2}=0.252$ ($24\farcs4$), in good agreement with the observed arcs.

Conversely, given a profile $(\alpha_0,p)$ and an Einstein radius $x_{\mathrm{c}}$ we can calculate a geometry factor $f$, solving \eqref{eq:critical} accordingly:
\begin{equation}\label{eq:plaw_f}
 f=\frac{1}{\alpha_0}x_{\mathrm{c}}^{1-p}.
\end{equation}
This enables us to test how a wrong choice of parameters affects the deduced geometry factor and compare the results to the likelihood peaks.
In the histograms in Fig. \ref{fig7}, the peak locations expected from the analytical estimation are marked along with the correct values. They agree remarkably well with the actual likelihood maxima. It seems that the shift in the distributions away from the true geometry factor can indeed be accounted for by an erroneous choice of mass profile parameters. Since such a choice does not change the features of the distributions, it should generally be difficult to distinguish it from other cases with better choices and remove the effect.

%% file: sec5.tex
\section{Semi-analytic calculation of geometry factors}\label{sec:semi}
We test a final approach that is based on the same parametric profile but aims to reproduce critical points rather than images. While it seems that arc configurations can often be predicted by a variety of lens models in different geometries, to some extent owing to the freedom in the source shapes and the position of images relative to the critical line, the shape of the critical curves itself may be more demanding to reproduce.

We consider a lens model with convergence $\kappa_0$ and shear $\gamma_0=\sqrt{\gamma_{0,1}^2+\gamma_{0,2}^2}$ for a source redshift $z_{\mathrm{r}}$. For a second source redshift $z_{\mathrm{s}}$ characterized by the geometry factor $f$ the convergence at any point $(x,y)$ in the lens plane has the value $\kappa(x,y)=f\kappa_0(x,y)$; the shear also scales with the geometry factor, $\gamma(x,y)=f\gamma_0(x,y)$. On the critical lines
\begin{equation}
 \mathrm{det}\,J(x,y)=[1-\kappa(x,y)]^2-\gamma^2(x,y)=0
\end{equation}
has to hold. Given a set of $N$ points $(x_{\mathrm{i}},y_{\mathrm{i}})$ known to lie on the critical line at redshift $z_{\mathrm{s}}$, we therefore take the function
\begin{equation}\label{eq:chi2}
 \chi^2=\frac{1}{N}\sum_{i=1}^N\,\left\{\left[1-f\kappa_0(x_{\mathrm{i}},y_{\mathrm{i}})\right]^2-f^2\gamma_0^2(x_{\mathrm{i}},y_{\mathrm{i}})\right\}^2
\end{equation}
to measure the agreement between the model and the data for the geometry factor $f$ and demand that its minimum should determine the best-fitting geometry factor. To apply this concept we first have to obtain sets of critical points. Ideally an estimator point is obtained by marking the brightness saddle point in an arc that appears to be formed by two merging images. In arcs exhibiting no such structure the brightest point can be chosen instead. It is useful to add more points nearby to mark the presumed direction of the critical line in this point since this makes for more stringent constraints. Multiple image systems can also be taken into account. If several arcs are observed at the same redshift at different positions in the lens plane, this translates into more information on the critical line and can help to constrain ellipticity in particular. 

We test the method using our mock observations of the cluster \textit{g1}. Since very few and only faint arcs are found at redshift $z=0.7$, we include only redshifts $z=1$, $z=2$ and $z=4$. Note that in principle we could fix the lens model for one of these redshifts by setting $f=1$ in Eq. \eqref{eq:chi2} and minimising with respect to the lens parameters. However it is all but impossible to determine a unique best-fitting model since a large number of combinations of lens parameters produces critical curves passing through our estimators, some with scale radii several times as large as the value suggested by other mass profile fits for this cluster. To avoid this obstacle, we rely on the same set of fixed lens parameters as for the MCMC method. We then use estimator sets for all three redshifts to compute their geometry factors. Table \ref{tab1} shows the values obtained in this way.

\begin{table}
\caption{Geometry factors $f$ computed for three source redshifts $z_{\mathrm{s}}$ from the chi-square minimisation, with a reference source redshift of either $z_{\mathrm{r}}=0.7$ or $z_{\mathrm{r}}=1.0$, and the theoretical $\Lambda$CDM values. For a reference source redshift of $z_{\mathrm{r}}=1.0$, the geometry factor for source redshift $z_{\mathrm{s}}=1.0$ is $f=1$ by definition.}\label{tab1}
\centering
 \begin{tabular}{cccc}
\hline\hline
$z_{\mathrm{r}}$ & $z_{\mathrm{s}}$ & $f$ (comp.) & $f$ ($\Lambda$CDM) \\
\hline
& 1.0 & 1.2334 & 1.2168 \\
0.7 & 2.0 & 1.5256 & 1.4634 \\
& 4.0 & 1.6058 & 1.5839 \\
\hline
1.0 & 2.0 & 1.2369 & 1.2027 \\
& 4.0 & 1.3019 & 1.3017 \\
\hline
 \end{tabular}
\end{table}

\begin{table*}
\caption{Influence of the scale radius $r_{\mathrm{s}}$ and ellipticity $\epsilon$ estimates on the geometry factor $f$ in the semi-analytical method. In the third column, the previous parameter choice and results are given. Results in the remaining columns were computed with a changed value of either $r_{\mathrm{s}}$ or $\epsilon$ as indicated in the header.}\label{tab2}
\centering
\begin{tabular}{ccccccc}
\hline\hline
$z_{\mathrm{r}}$ & $z_{\mathrm{s}}$ & $r_{\mathrm{s}}=0.308\,h^{-1}\,\mathrm{Mpc}$, $\epsilon=0.37$ & $r_{\mathrm{s}}=0.317\,h^{-1}\,\mathrm{Mpc}$ & $r_{\mathrm{s}}=0.294\,h^{-1}\,\mathrm{Mpc}$ & $\epsilon=0.35$ & $\epsilon=0.40$ \\
\hline
& 1.0 & 1.2334 & 1.2150 & 1.2631 & 1.2707 & 1.1782 \\
0.7 & 2.0 & 1.5256 & 1.4991 & 1.5685 & 1.5800 & 1.4452 \\
& 4.0 & 1.6058 & 1.5771 & 1.6522 & 1.6573 & 1.5297 \\
\hline
1.0 & 2.0 & 1.2369 & 1.2338 & 1.2418 & 1.2434 & 1.2266 \\
& 4.0 & 1.3019 & 1.2980 & 1.3081 & 1.3042 & 1.2983 \\
\hline
 \end{tabular}
\end{table*}
Two different reference source redshifts are listed for the following reason: The value for the scale convergence $\kappa_{\mathrm{s}}$ is originally defined here for a source redshift of $z_{\mathrm{r}}=0.7$. If we are confident that the model is valid for this redshift, we can consider the geometry factors with respect to the same reference redshift. If we do not trust the model however, we can restrict our analysis to the redshifts that actually appear in observations. Since
\begin{equation}
 f\left(z_{\mathrm{s}}=z_1,z_{\mathrm{r}}=z_2\right)=\frac{f\left(z_{\mathrm{s}}=z_1,z_{\mathrm{r}}=z_0\right)}{f\left(z_{\mathrm{s}}=z_2,z_{\mathrm{r}}=z_0\right)}
\end{equation}
for arbitrary redshifts $z_{0,1,2}$ according to the definition of the geometry factor (cf. Eq. \eqref{eq:f}), we consider ratios between the computed geometry factors. While for instance $f\left(z_{\mathrm{s}}=2.0,z_{\mathrm{r}}=0.7\right)$ is the factor by which the input scale convergence has to be multiplied for the critical line to match our estimator set for redshift $z_{\mathrm{s}}=2$, $f\left(z_{\mathrm{s}}=2.0,z_{\mathrm{r}}=1.0\right)$ instead can be taken to quantify the scaling between our two estimator sets. In other words, the convergence is rescaled to reproduce the critical line at redshift $z_{\mathrm{s}}=1.0$ and geometry factors for higher redshifts are considered with respect to that redshift. This has the advantage of eliminating the influence of the scale convergence from our calculations. Naturally, the remaining lens parameters can still act as sources of errors.

Table \ref{tab1} confirms these considerations to some degree: for the reference source redshift of the lens parameters, $z_{\mathrm{r}}=0.7$, deviations of the geometry factors are roughly 4 per cent for a source redshift of $z_{\mathrm{s}}=2$ and  1 per cent for $z_{\mathrm{s}}=4$, whereas considering the ratios they reduce to 3 per cent and less than 0.1 per cent respectively. The excellent agreement in the latter case is certainly to some extent coincidental. Taking our lack of knowledge of the precise mass model into account, uncertainties should be considerably larger than the deviation itself.

To find out how our parameter choice affects the results, we simply rerun the program for different lens models. Having just described how to eliminate the scale convergence, we investigate the role of the scale radius $r_{\mathrm{s}}$ and the ellipticity $\epsilon$. Table \ref{tab2} lists the geometry factors computed for various cases in which changed values for either the scale radius or the ellipticity were used.

For a reference source redshift of $z_{\mathrm{r}}=0.7$, the geometry factor results vary over a range corresponding to about 8 per cent across our examples of lens models. Changing the reference redshift - that is, again considering only ratios - the scatter reduces to 1 per cent. In that case, however, all results calculated for $z_{\mathrm{s}}=2$ overestimate the geometry factor without exception, due to errors in either the assumed mass model or our choice of critical line estimators. It is also worth noting that in this example we used multiple arcs at the same redshift to derive estimators. In practice, fewer and fainter arcs, albeit at more redshifts, might be available, complicating the choice of reliable estimators.

%% file: sec6.tex
\section{Conclusions}\label{sec:conclusions}
We have investigated several approaches to cluster strong lensing to test its use as a probe of spacetime geometry, in particular the dark energy equation of state. Generally this has proven challenging as changes in the observables induced by variations in the equation of state are very small. The methods that we have explored make use of the growth of the critical lines with the source redshift. We have tried to constrain a cosmologically sensitive ratio of angular diameter distances, using first a Metropolis-Hastings algorithm constructed to fit an observed image configuration and secondly a semi-analytic approach that optimizes the distance ratio such that a fixed lens model reproduces a given set of critical points. To test our methods we have studied strong lensing by a numerically simulated cluster, for which we have created mock observations.

Modelling uncertainties have turned out to be the most important source of errors in our study. The simplest models for the lensing mass distributions, such as the SIS and the NFW profile, assume spherical symmetry. They offer the advantage of admitting a largely analytic analysis and thus help us estimate geometric effects, but they are generally not well suited to a study of lensing by a real galaxy cluster as in most cases deviations from the assumed symmetry are too large. 

Models that include ellipticity provide a more accurate description of the mass distributions. The larger number of parameters, however, produces degeneracies. If all model parameters are treated as free, any cosmological sensitivity is masked since the background geometry and profile properties can influence the lensing behaviour in a similar way. An attempt to optimise the lens along with the geometry for an individual cluster does not result in meaningful constraints on the distance ratio, but instead admits a wide range of realisations of dark energy. Constraints can be narrowed down somewhat if the lens model is fixed and only the geometry is optimised, but care must be taken in the choice of the model. While the influence of the overall mass assumed is weak if we choose to consider only the scaling of the critical curve between two redshifts, the remaining parameters can still create biases. Ideally, no geometric assumptions should be made to fix the lens model, yet information from arcs at a single redshift alone is not sufficient to constrain it. Independent information from other observations could be used to address the problem, but it is generally difficult to obtain for the cluster core. \citet{2005MNRAS.362.1301M} suggested that the position of the brightest cluster galaxy (when present) or the centre of X-ray emission could be referred to to determine the cluster centre, but they pointed out that errors of several arcseconds can occur. Additional constraints could be derived from the galaxy velocity dispersion, weak gravitational lensing or X-ray temperature profiles.

It is also worth investigating whether the restriction to analytical profiles can contribute significant errors. Galaxy clusters are generally 'lumpy', which gives rise to the question if an ellipsoidal NFW profile nonetheless describes the lensing properties well enough to permit cosmological conclusions. If even the best fitting NFW profile is insufficient, biases are to be expected. An interesting approach to cluster strong lensing that forgoes the use of the NFW profile was presented by \citet{2009MNRAS.396.1985Z}. Assuming that mass follows light in a cluster, they assigned a power law profile to each visible cluster galaxy, scaled by the observed brightness, and smoothed the resulting distribution. Note that they based their method on multiple images. While we have not considered this effect in our study, it provides additional constraints and should therefore be a useful inclusion.